      [1999/12/01 v1.4c Il Nuovo Cimento]
\documentclass{cimento}

\usepackage{graphicx}  

\def\MS{\hbox{$\overline{\rm MS}$}}

\title{Top production at the LHC: the impact of
PDF uncertainties and correlations}

\author{Alberto~Guffanti\from{ins:Freiburg} and
Juan~Rojo\from{ins:Milano}
}
\instlist{
  \inst{ins:Freiburg} Physikalisches Institut, Albert-Ludwigs-Universit\"at,
  Hermann-Herder-Str. 3, 79104, Freiburg i. Br., Germany.
  \inst{ins:Milano} Universit\`a degli Studi \& INFN Sezione di Milano, 
  via Celoria 16, 20100 Milano, Italy.}
\PACSes{\PACSit{12.38t}{Quantum chromodynamics}
        \PACSit{14.65.Ha}{Top quarks}}

\begin{document}

\maketitle

\begin{abstract}
In this contribution we discuss the impact of PDF uncertainties
on $t\bar{t}$ and $t$-channel single-top 
production at the LHC. We present predictions for total 
cross-sections computed at NLO accuracy with different PDF sets.
For single-top production, we point out that the uncertainty arising
from the choice of the bottom quark mass is one of the dominant
theoretical uncertainties on the total cross-section.
Finally, the possibility of using PDF induced correlations between 
top quark and
electroweak vector boson  production 
cross-sections to improve the accuracy of LHC measurements is 
investigated. 

\end{abstract}

\section{Introduction}
A precise determination of Parton Distribution Functions (PDFs) with 
reliable estimate of their uncertainties is crucial for the success of 
the physics program at the LHC experiments. 
On the one hand PDF uncertainties are often the dominant theoretical 
uncertainties for many relevant signal and background 
processes~\cite{Campbell:2006wx}. 
On the other hand overestimated PDF errors might hinder 
the discovery of new physics effects, as shown for example 
in~\cite{Ferrag:2004ca}.
Top physics at the LHC is no exception and both top pair and single-top
production present complementary and interesting properties as far as PDF
determination/effects are concerned. This contribution aims to review
some of the implications of PDFs for top quark physics at the LHC.

In the first part of the contribution we summarize the present 
status of the predictions for $t\bar{t}$, $t-$ and $s-$channel single-top 
cross-sections, computed using different PDF sets. We show how differences 
in the predictions can directly be traced to both differences in the parton 
luminosities, and the values of physical parameters used in the PDF 
analyses, such as the strong coupling constant $\alpha_s$ or the  
the $b$-quark mass, $m_b$.
In particular, we highlight the importance to account for the uncertainty 
on the $b$-quark mass for accurate predictions of single-top production at 
the LHC.

In the second part we study PDF induced correlations between PDFs and 
$t\bar{t}$/single-top cross-sections and between top and $W^\pm$/$Z^0$ 
cross-sections. 
We briefly discuss how these correlations could be used in order to 
improve the accuracy of top cross-section measurements with early data 
at the LHC.
These correlation studies are performed within the framework of the 
NNPDF parton analysis~\cite{DelDebbio:2007ee,Ball:2008by,Ball:2009mk,
Ball:2010de}  which, by relying on Monte Carlo techniques 
for the estimation of uncertainties, provides an ideal tool for such 
statistical studies.

The baseline PDF set for the studies presented in this contribution
is the recently released NNPDF2.0~\cite{Ball:2010de}, the first NLO global 
fit using the NNPDF methodology. 

\section{Top-quark production at the LHC}

\subsection{$t\bar{t}$ production}

Top pair production is the main channel for top quark production at Tevatron 
and LHC.
In Table~\ref{tab:ttbar} we collect the predictions for the top pair
cross-section at LHC 7 TeV at NLO computed with the MCFM code~\cite{ref:MCFM} 
using different PDF sets.

\begin{table}[ht!]
  \caption{Top pair cross-section at NLO with different PDF sets at 
    LHC 7 TeV.}
  \label{tab:ttbar}
  \begin{narrowtabular}{2cm}{c|c}
    \hline
      CTEQ6.6~\cite{Nadolsky:2008zw}  & 147.7 $\pm$ 6.4 pb \\
      MSTW2008~\cite{Martin:2009iq}   & 159.0 $\pm$ 4.7 pb \\
      NNPDF2.0~\cite{Ball:2010de}     & 160.0 $\pm$ 5.9 pb \\
    \hline
      ABKM09~\cite{Alekhin:2009ni}    & 131.9 $\pm$ 4.8 pb \\
      HERAPDF1.0~\cite{:2009wt}       & 136.4 $\pm$ 4.7 pb \\
    \hline
  \end{narrowtabular}
\end{table}

We notice that the predictions from the three global fits, NNPDF2.0, CTEQ6.6 
and MSTW08 agree at the 1-sigma level.
The differences with PDF sets based on reduced datasets, ABKM09 and HERAPDF1.0, 
are larger. 
We note that top pair production depends strongly on the large-$x$ gluon, 
and thus using sets which do not include Tevatron jet data might lead to rather 
different predictions for this observable.
One should notice that differences between the predictions from different PDF 
sets also arise from the use of different values for the strong coupling 
constant $\alpha_s$.  
It has been shown that using a common value of $\alpha_s$ brings predictions
from different groups for various LHC observables, including top pair 
production, in better agreement~\cite{Demartin:2010er,Ubiali:2010xc}.

\begin{figure}[ht!]
  \centering
  \includegraphics[width=7cm]{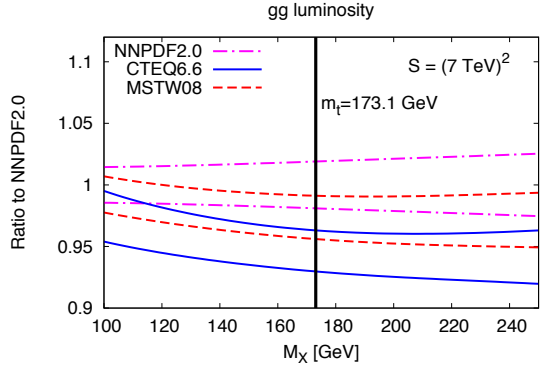}
  \caption{Gluon-gluon parton luminosities for CTEQ6.6, MSTW2008 and NNPDF2.0
    including the associated PDF uncertainties. Results are showed as ratios
    to NNPDF2.0.}
  \label{fig:gglumi}
\end{figure}

Once we subtract the difference introduced by different choices for $\alpha_s$, 
the remaining differences can be directly traced to differences in the PDF 
luminosities at the typical scale of the process.
This is illustrated in Fig.~\ref{fig:gglumi}, where the gluon-gluon luminosity
for LHC at 7 TeV is plotted for the CTEQ6.6, MSTW2008 and NNPDF2.0 NLO sets.
For example, the lower value for the cross-section obtained using the CTEQ6.6
set reflects the smaller $gg$ luminosity as compared to the other sets at
$Q^2=m_t^2$.

\subsection{Single-top production}

Next we present predictions for single-top production at the LHC at 7 TeV.
These predictions for various PDF sets are collected in 
Table~\ref{tab:singletop}, where we present results for both the $t$- 
and $s$-channel single-top cross-sections computed at NLO in QCD with 
the MCFM code. We have used the $N_f=5$ (massless) calculation 
for the results presented here, recently the $N_f=4$ calculation, which
properly takes into account the effects due to finite $b$-quark mass, also 
became available~\cite{Campbell:2009ss}.

While at the Tevatron the contributions of $t$- and $s$-channel $W$ exchange 
to single-top production are comparable in size, at the LHC $t$-channel 
production is by far the dominant production mechanism.

From the point of view of testing the predictions from different PDF sets 
$t$-channel single-top is also very interesting due to the fact that, in the 
so called 5-flavour scheme ({\it i.e.} a scheme where the $b$ is assumed to
be a parton in the proton) the cross-section at LO probes directly the 
$b$-quark PDF, which in turn is closely related to the gluon distribution 
from which it is generated radiatively.

From Table~\ref{tab:singletop} one notices that the central predictions from 
the various PDF sets can differ by several times the quoted 1-sigma PDF 
uncertainty\footnote{Note that for the ABKM09 prediction the uncertainty 
includes the associated $\alpha_s$, $m_c$ and $m_b$ uncertainties, which 
cannot be disentangled from the PDF uncertainties, and is thus much larger 
than the one obtained using other PDF sets.}.
There are different contributions to this discrepancy. The first stems from 
the different values of the strong coupling constant $\alpha_S$ which are 
used by different parton sets. Since single-top production is mediated
by electroweak gauge bosons, $\alpha_s$ enters only in radiative corrections,
unlike the case of $t\bar{t}$ production discussed above, this
effect is rather small.

\begin{figure}[ht!]
  \centering
  \includegraphics[width=0.45\textwidth]{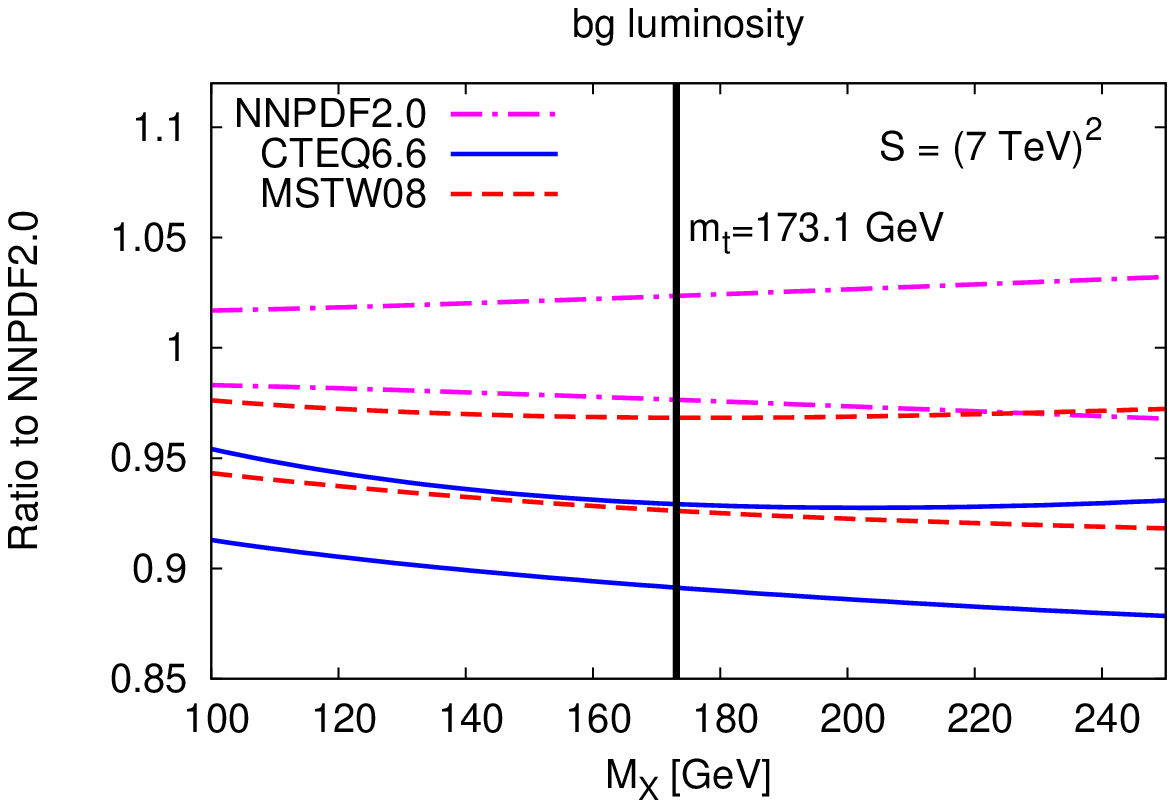}\qquad
  \includegraphics[width=0.45\textwidth]{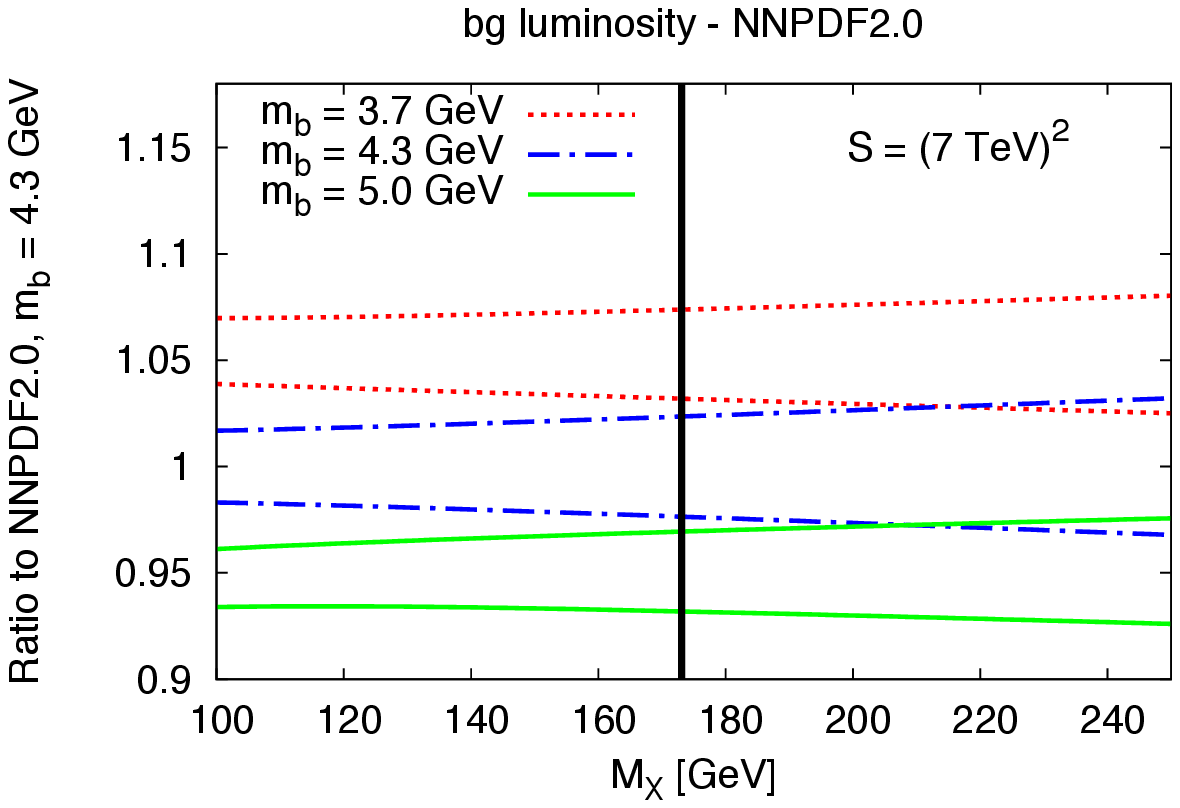}
  \caption{(left) $b$-gluon parton luminosities for CTEQ6.6, MSTW2008 and 
    NNPDF2.0, normalized to the NNPDF2.0 value. (right) $b$-gluon parton 
    luminosities for NNPDF2.0 fits with different values of the $b$-quark 
    mass normalized to the standard NNPDF2.0.}
  \label{fig:bglumi}
\end{figure}

In order to separate the differences in the single-top production 
cross-section which arise from the differences in the PDFs themselves
and those from other physical parameters (like $m_b$ or $\alpha_s$) which 
also enter in the PDF analyses and in the computation of the partonic matrix 
element, we plot in Fig.~\ref{fig:bglumi} the $b$-gluon parton luminosity, 
which determines the LO cross-section, for the CTEQ6.6, MSTW2008 and 
NNPDF2.0 NLO sets. 
It is clear that parton luminosities in the kinematic region relevant for 
single-top production differ by an amount much smaller than the cross-sections 
themselves, suggesting that the differences indeed come from variations of 
other physical parameters which enter the PDF analysis.

Indeed, it can be seen that the bulk of this difference is related to the 
different values of the $b$-quark mass used in the fits by the different
collaborations. The NNPDF collaboration sets $m_b=4.3$ GeV, CTEQ uses 
$m_b=4.5$ GeV while the MSTW08 fit is performed setting $m_b=4.75$ GeV. 
In order to substantiate our claim that the different values used the $b$-quark 
mass explain the bulk of the difference for the $t$-channel single-top 
cross-section, we produced two NNPDF2.0 sets with $m_b = 3.7$ GeV 
and  $m_b = 5.0$ GeV respectively. 
The results for the $b$-gluon parton luminosities for these modified 
sets are shown in the right plot in Fig.~\ref{fig:bglumi} and the 
corresponding cross-sections for the $t$-channel single-top cross-section 
are collected in Table~\ref{tab:singletop-mb}.
It is clear that the value of $m_b$ is anti-correlated with the $bg$ luminosity 
and the $t$-channel single-top cross-section.

From Table~\ref{tab:singletop-mb} one sees that variations of the $b$-quark 
mass of the order or $\delta m_b\sim 0.7$ GeV induce an uncertainty in the 
cross-section of $\delta\sigma \sim 3$ pb.
If we take the PDG average as the best available determination of the $b$-quark
mass and convert it from the \MS scheme to the pole mass scheme we obtain an
uncertainty of approximately $\delta m_b({\rm PDG})\sim 0.2$ GeV. Rescaling 
errors, we are still left with an uncertainty in the $t$-channel single-top 
cross-section of $\delta\sigma \sim 0.8$ pb, still larger than the typical 
nominal PDF uncertainties quoted in Table~\ref{tab:singletop}. 
It is also clear from Table~\ref{tab:singletop-mb} and Fig.~\ref{fig:bglumi} 
that using a similar value of $m_b$ would bring the predictions from different 
PDF sets into much better agreement. The uncertainty due to $m_b$ should 
thus always be accounted for in the theoretical predictions for LHC single-top 
production.

\begin{table}[t!]
  \caption{Single-top cross-section at NLO with different PDF sets at 
    LHC 7 TeV.}
  \label{tab:singletop}
  \begin{narrowtabular}{2cm}{c|c|c}
    \hline
                                    & $t$-channel            & $s$-channel   \\ 
    \hline
    CTEQ6.6~\cite{Nadolsky:2008zw}  &  40.85 $\pm$  0.50 pb & 2.33 $\pm$ 0.05 pb\\
    MSTW2008~\cite{Martin:2009iq}   &  41.96 $\pm$  0.26 pb & 2.38 $\pm$ 0.04 pb\\
    NNPDF2.0~\cite{Ball:2010de}     &  44.33 $\pm$  0.32 pb & 2.38 $\pm$ 0.06 pb\\
    \hline
    ABKM09~\cite{Alekhin:2009ni}    &  43.17 $\pm$  1.98 pb & 2.40 $\pm$ 0.03 pb\\
    HERAPDF1.0~\cite{:2009wt}       &  40.04 $\pm$  0.33 pb & 2.38 $\pm$ 0.05 pb\\
    \hline
  \end{narrowtabular}
\end{table} 

\begin{table}[t!]
  \caption{$t$-channel single-top cross-section at NLO computed using NNPDF2.0 
    sets with different values of the $b$-quark mass.}
  \label{tab:singletop-mb}
  \begin{narrowtabular}{2cm}{l|c}
    \hline
    NNPDF2.0 ($m_b=3.7$ GeV) &  46.77 $\pm$ 0.36 pb \\
    NNPDF2.0 ($m_b=4.3$ GeV) &  44.33 $\pm$ 0.32 pb \\
    NNPDF2.0 ($m_b=5.0$ GeV) &  41.04 $\pm$ 0.32 pb \\
    \hline
  \end{narrowtabular}
\end{table}

\section{PDF-induced correlations}

It is well known (see for example the discussion in 
Ref.~\cite{Nadolsky:2008zw,Ball:2008by}) that parton densities
induce correlations among different observables measured at hadron 
colliders. 
These can be the PDFs themselves, one PDF and a physical observable or 
two physical observables. The latter case is especially important from 
the experimental point of view, since it allows to define measurement 
strategies in which the PDF uncertainties between two observables cancel, 
for example in the case in which these correlation between the two observables 
is maximal.

In the case of a PDF set based on the Monte Carlo method, like NNPDF, 
the correlation coefficient $\rho[A,B]$ for two observables $A$ and $B$ which 
depend on PDFs is given by the standard expression for the correlation of two
stochastic variables~\cite{Ball:2008by,Demartin:2010er}
\begin{equation}
  \label{eq:correlation}
  \rho[A,B]=\frac{\langle A B\rangle_{\mathrm{rep}}
    - \langle A\rangle_{\mathrm{rep}}\langle B\rangle_{\mathrm{rep}} }
  {\sigma_A\sigma_B}
\end{equation}
where the averages are taken over the ensemble of the $N_{\mathrm{rep}}$ values 
of the observables computed with the different replicas of the PDF set, 
and $\sigma_{A,B}$ are the standard deviations for the observables as computed 
from the MC ensemble.
The value of $\rho$ characterizes whether two observables are correlated 
($\rho \approx 1$), anti-correlated ($\rho \approx -1$) or uncorrelated 
($\rho\approx 0$). In the following we present results for the NNPDF2.0 set, 
the LHC cross-sections have been obtained as before using the MCFM code.

\begin{figure}[ht!]
  \centering
  \includegraphics[width=0.33\textwidth,angle=270]{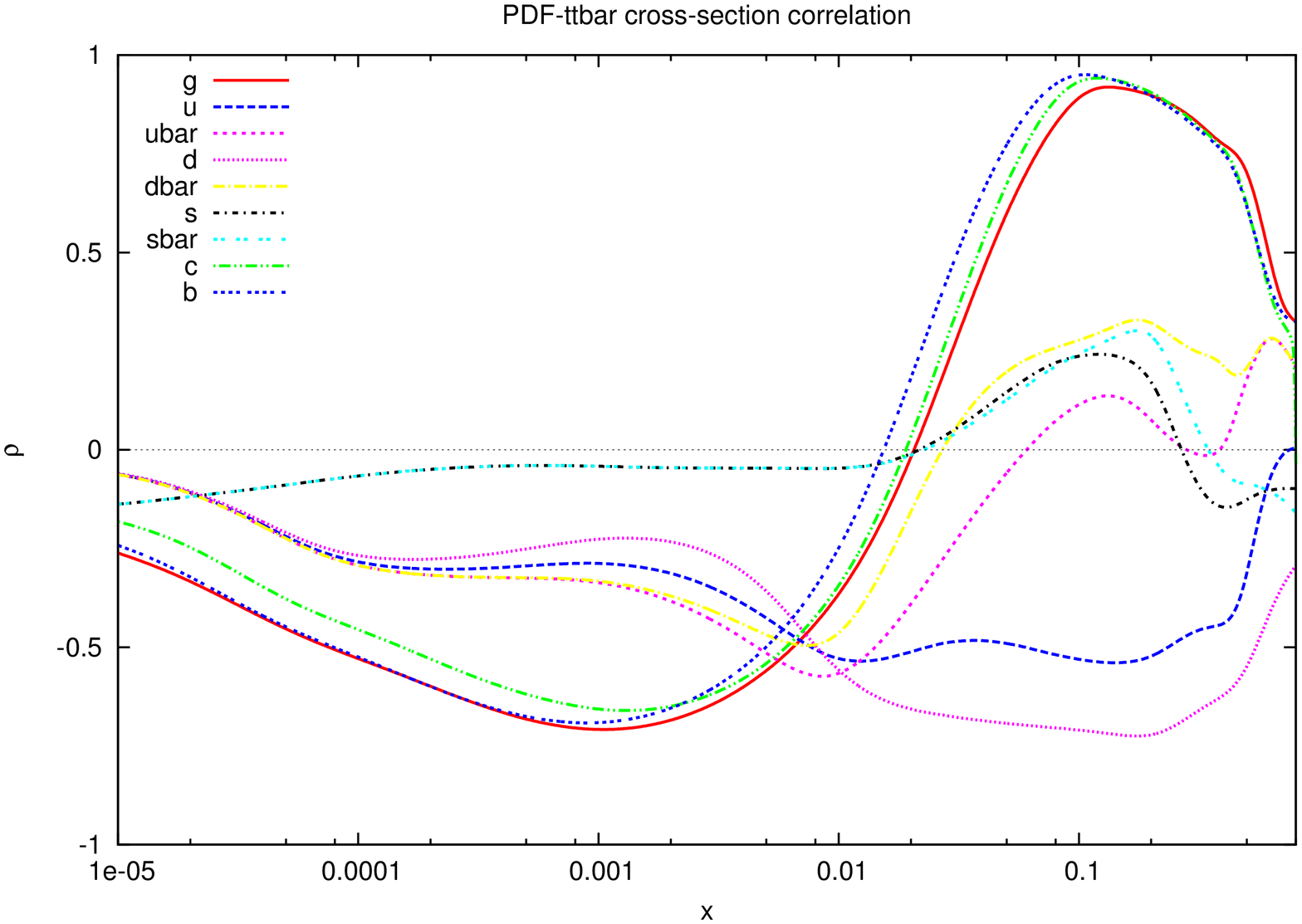}\qquad
  \includegraphics[width=0.33\textwidth,angle=270]{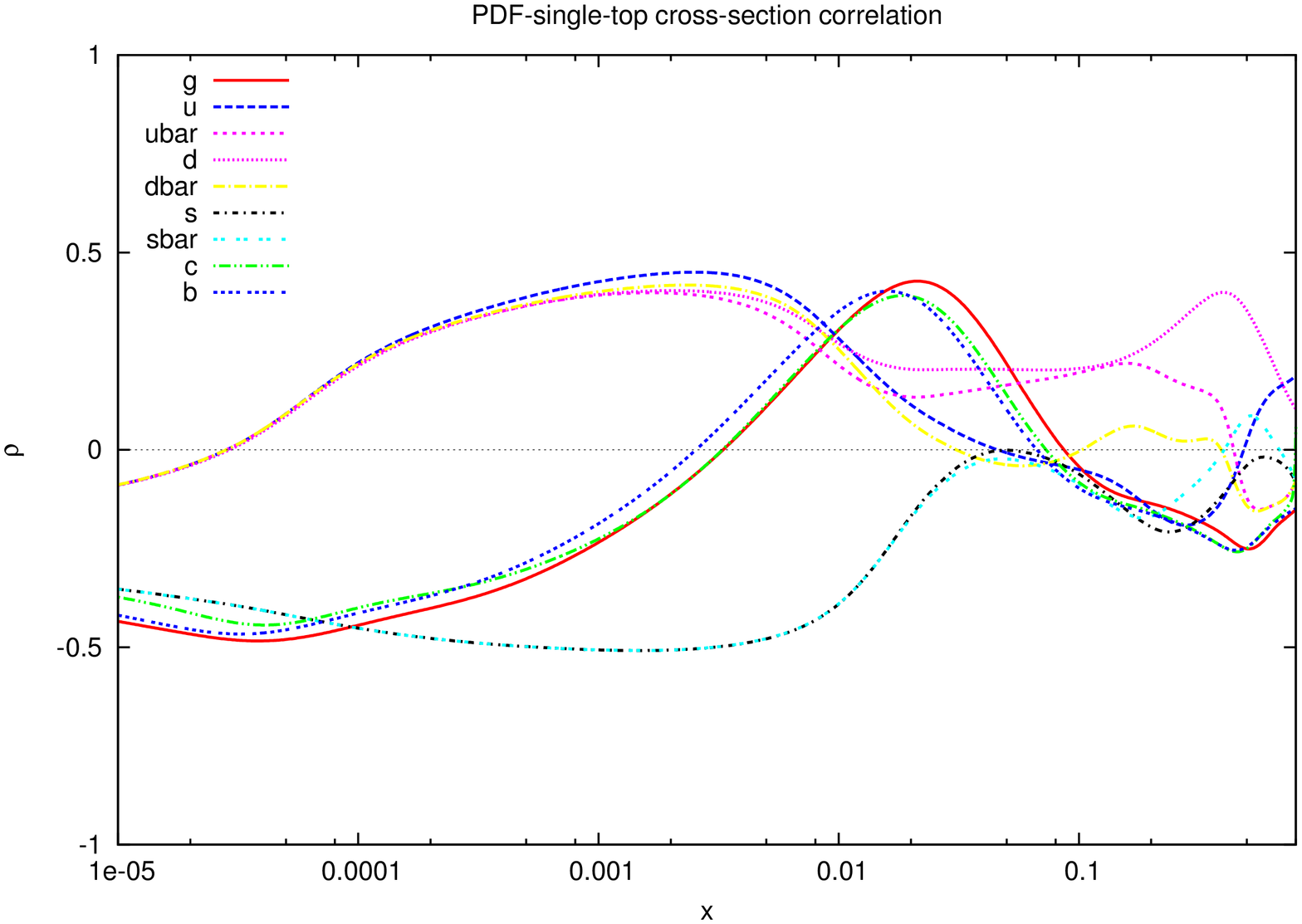}
  \caption{Correlation between parton densities and $t\bar{t}$ (left) 
    and $t$-channel single-top (right) cross-sections at the LHC 7 TeV. 
    The PDF set used is NNPDF2.0 and cross-sections have been
    computed with MCFM.}
  \label{fig:pdf_obs_corr}
\end{figure}

As a first example, we compute the correlation between the $t\bar{t}$ and
$t$-channel single-top cross-section at the LHC (7 TeV) and different PDFs
at the factorization scale $\mu_f=m_{\mathrm{top}}$ as a function of $x$.
The results are plotted in Fig.~\ref{fig:pdf_obs_corr}. 
The most remarkable features are that the $t\bar{t}$ cross-section at the 
LHC at 7 TeV is mostly correlated to the gluon distribution at $x\sim 0.1$ 
and anti-correlated with it at small-$x$, with the same behaviour present 
for sea quark PDFs,  generated radiatively from the gluon. 
We note also that the $u$- and $d$-quark distributions are anti-correlated 
with the cross-section at medium-/large-$x$. 

As for the case of the $t$-channel single-top cross-section, we point out 
that the strong correlation with the gluon (and therefore with the $c$- 
and $b$-quark distributions) present for $t\bar{t}$ is now milder 
and peaked at medium-$x$, $x\sim 0.01$, and now we find a moderate 
correlation at medium-/small-$x$ with the $u$ and $d$ PDFs, of the 
opposite sign as in the $t\bar{t}$ cross-section. 
We would like to stress as well the correlation of single-top
cross-section and the $s$-,$\bar{s}$-quark PDFs at medium-/small-$x$, which 
is notably absent in the $t\bar{t}$ case.

\begin{figure}[t!]
  \centering
  \includegraphics[width=0.31\textwidth,angle=90]{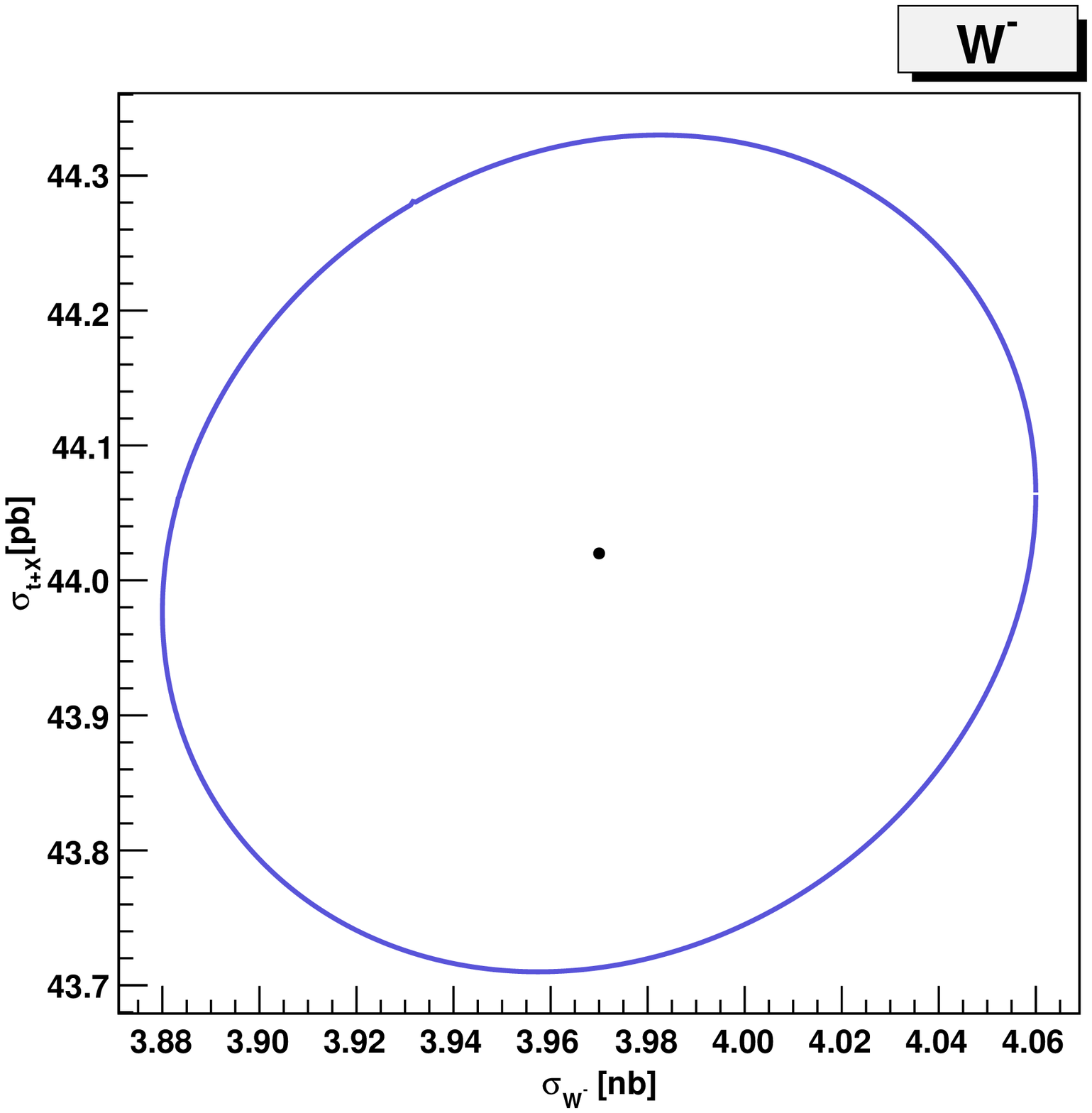}\quad
  \includegraphics[width=0.31\textwidth]{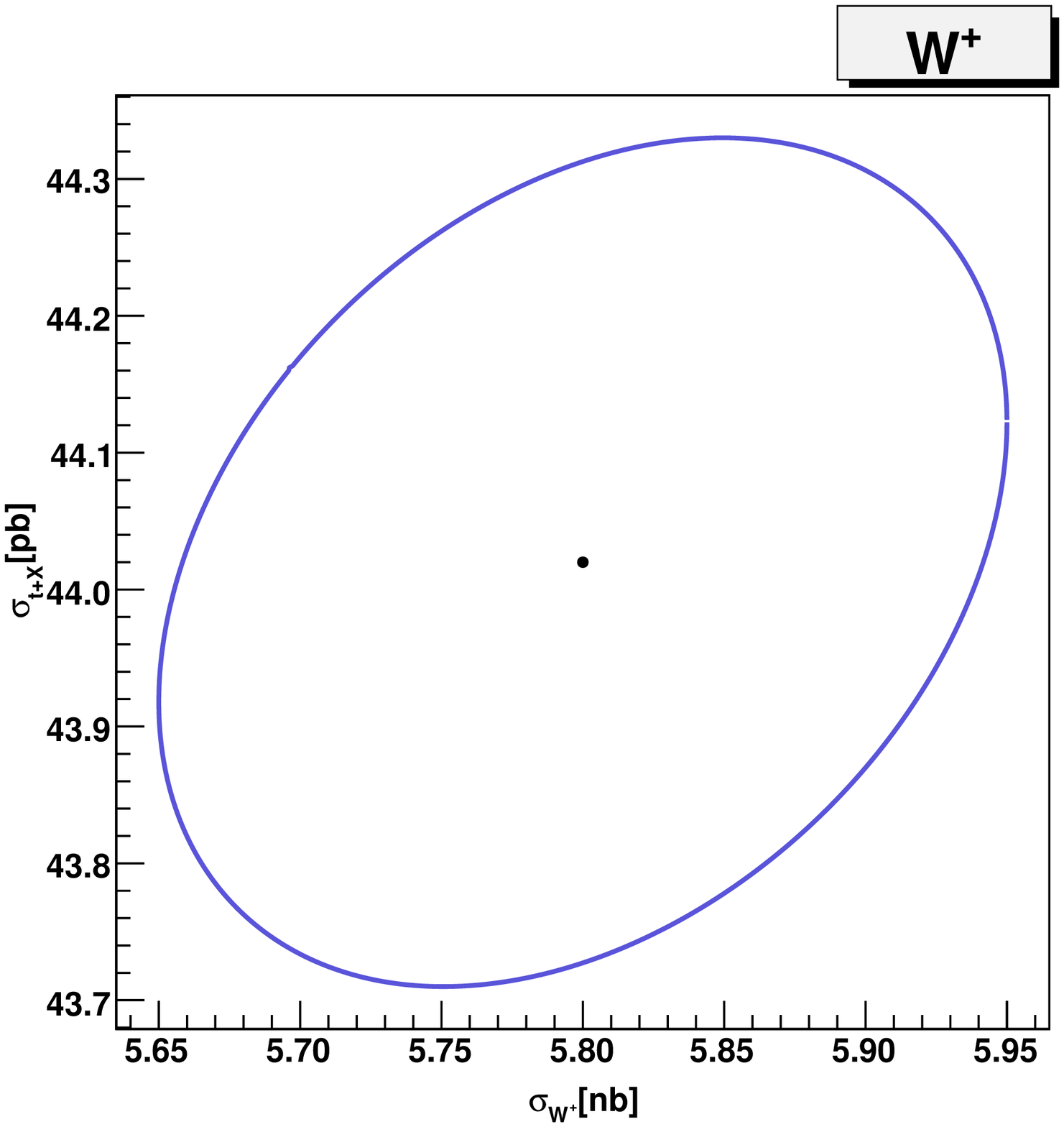}\quad
  \includegraphics[width=0.31\textwidth]{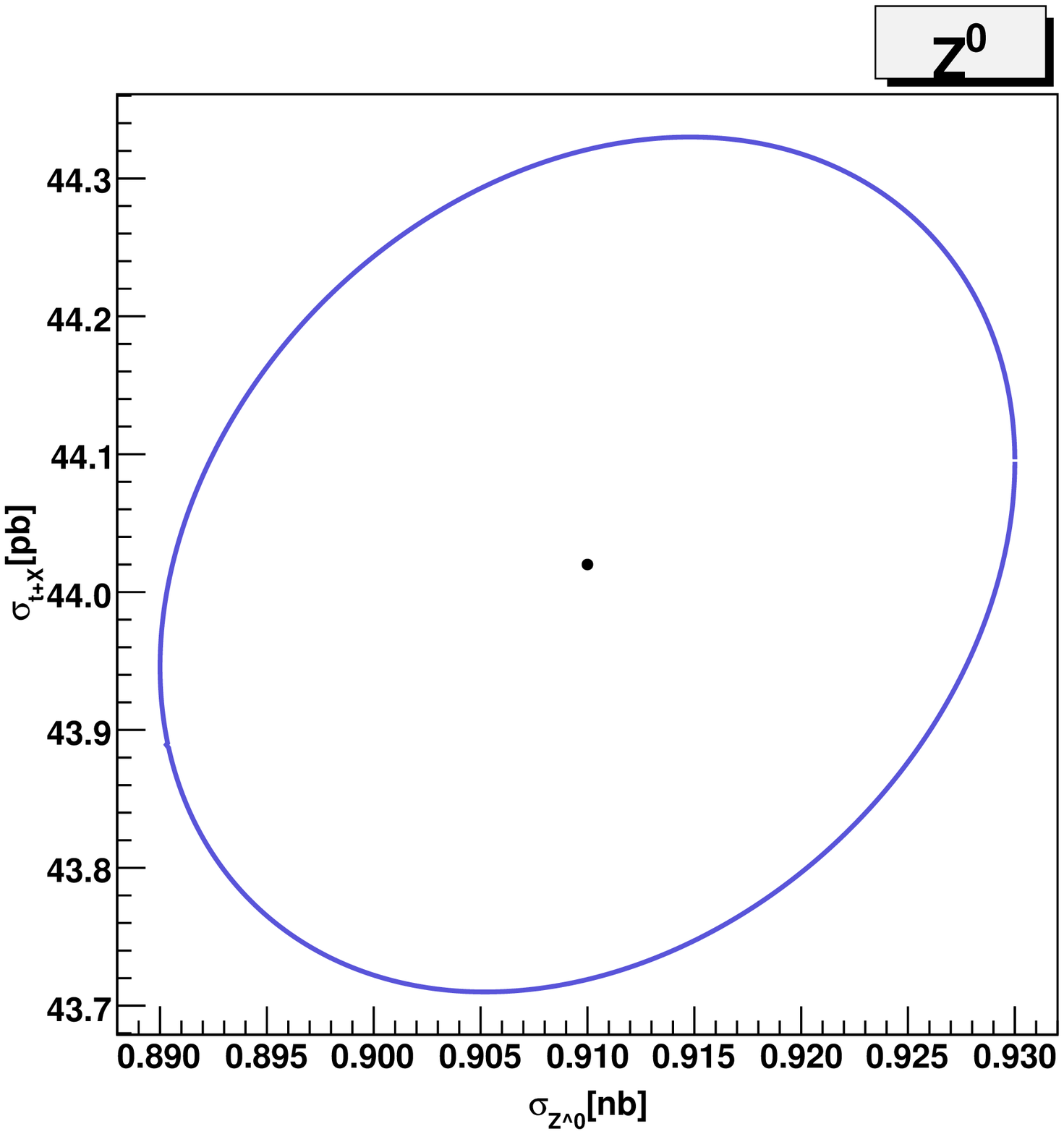}\quad
  \caption{Correlation between $t$-channel single-top and Electroweak Vector 
    Boson cross-sections at the LHC 7 TeV.  The cross-sections have
 been computed
with MCFM and the NNPDF2.0 parton set.}
  \label{fig:singletop_VB_corr}
\end{figure}

As previously pointed out, the correlation coefficient 
Eq.~(\ref{eq:correlation}) can also be computed between two cross-sections, 
which is potentially relevant since in the case of a sizable correlation 
the measurement of one of these observables would provide useful information 
on the value of the other one. In this respect we have computed the 
correlation between the $t\bar{t}$ and $t$-channel single-top cross-sections 
on one side and $W^{\pm}$ or $Z^0$ cross-sections at the LHC on the other.
The values for the correlation coefficients for the different pairs of 
observables are collected in Table~\ref{tab:obs_obs_corr} and the correlation 
ellipses are plotted in Fig.~\ref{tab:obs_obs_corr} for the $t\bar{t}$ 
cross-section and in Fig.~\ref{fig:singletop_VB_corr} for $t$-channel 
single-top.

\begin{table}[t!]
  \centering
  \begin{narrowtabular}{3cm}{c|c|c|c}
    \hline
    $\mathbf{\rho}$     & $\sigma_{W^+}$ & $\sigma_{W^-}$ & $\sigma_{Z^0}$\\
    \hline
    $\sigma_{t\bar{t}}$ &   -0.716       &    -0.694      &   -0.773      \\
    \hline
    $\sigma_{t}$        &    0.330       &     0.140      &    0.240      \\
    \hline
  \end{narrowtabular}
  \caption{Correlation coefficients between $t\bar{t}$ or $t$-channel 
    single-top and $W^{\pm}$ or $Z^0$ cross-sections at the LHC 7 TeV in the
    NNPDF2.0 analysis. }
  \label{tab:obs_obs_corr}
\end{table}

Both the values of $\rho$ and the shape of the correlation ellipses show 
a significant anti-correlation between the $t\bar{t}$ cross-section and 
the $W^\pm$ and $Z^0$ ones.
Given the fact that the vector boson cross-sections at the LHC are 
${\cal{O}}(10)$ times larger than the $t\bar{t}$ cross-section and 
more accurately known from the theoretical point of view, it is foreseeable 
to use those in order to better calibrate the top pair cross-section 
measurement in early data.

On the other hand the single-top cross-section shows a very mild 
correlation to the vector boson one thus rendering a similar approach based 
on the precision measurement of EW bosons difficult. However, if one is able
to identify other observables which should be measured with a similar
precision and that are correlated to the single-top cross-section, the
discussion of the $t\bar{t}$ case would also apply here.
 
\begin{figure}[t!]
  \centering
  \includegraphics[width=0.31\textwidth]{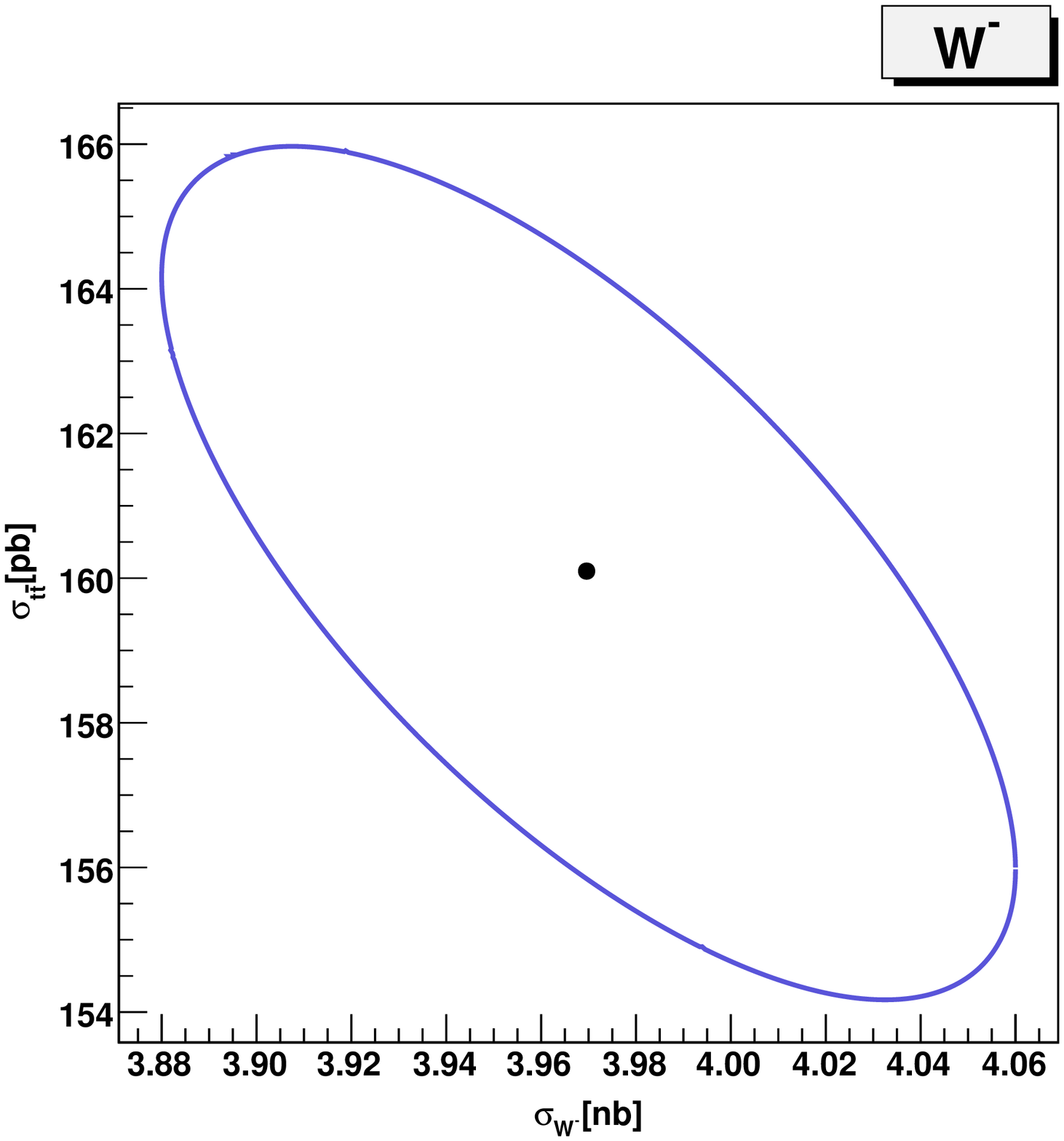}\quad
  \includegraphics[width=0.31\textwidth]{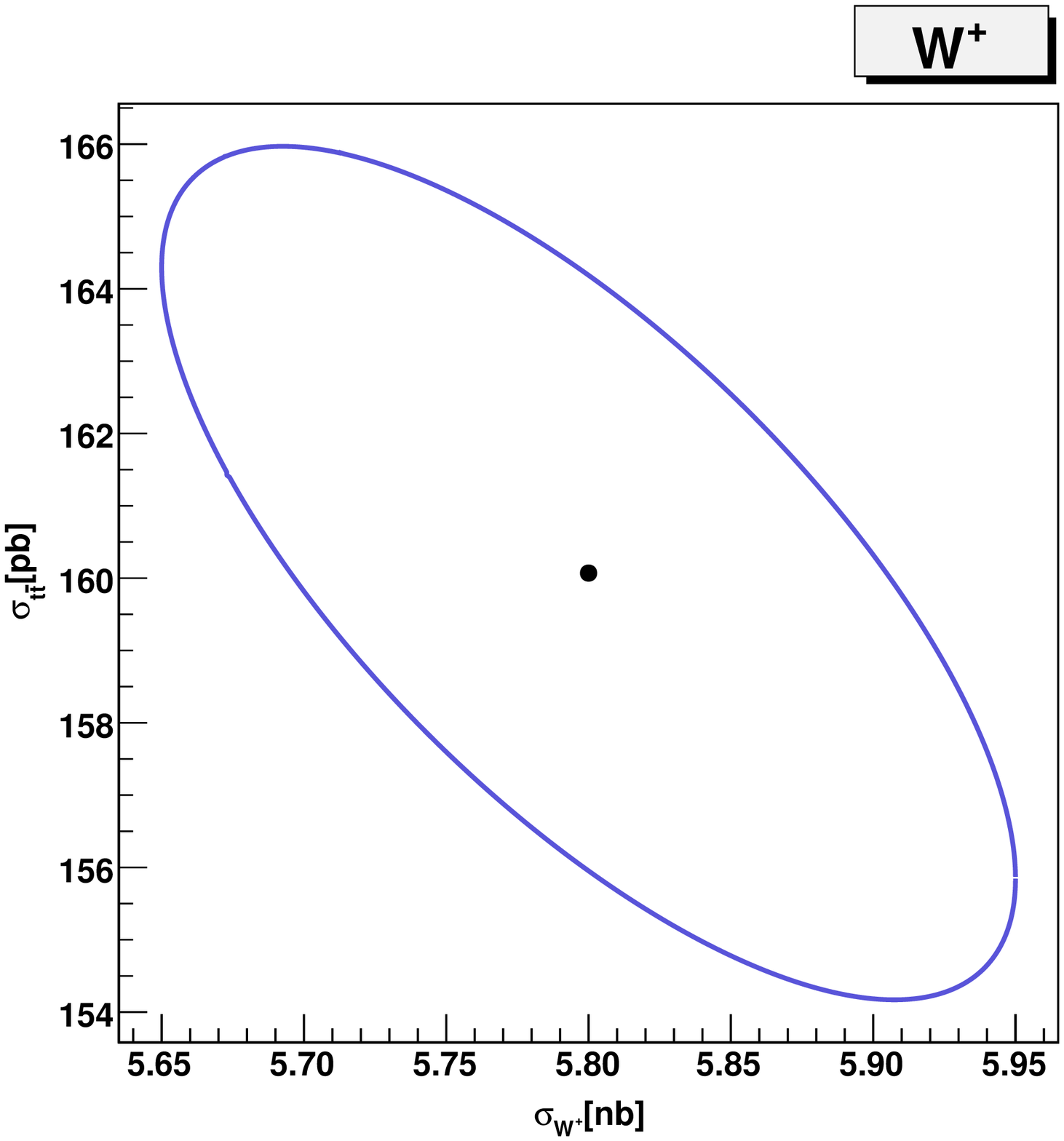}\quad
  \includegraphics[width=0.31\textwidth]{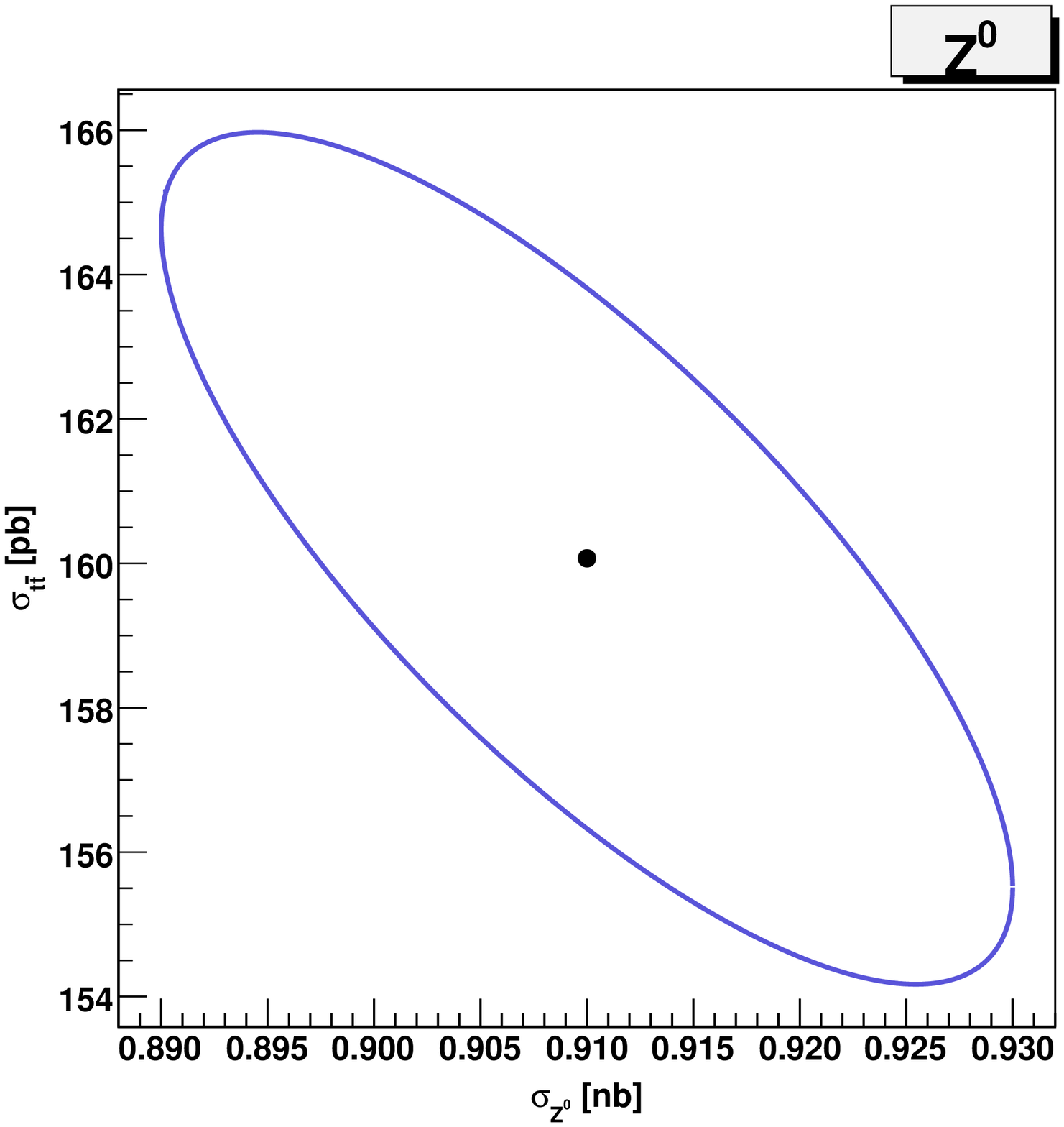}\quad
  \caption{Correlation between $t\bar{t}$ and Electroweak Vector Boson cross 
    sections at the LHC 7 TeV. The cross-sections have been computed
with MCFM and the NNPDF2.0 parton set.}
  \label{fig:ttbar_VB_corr}
\end{figure}

\section{Conclusions}

The quality of top physics results from the LHC experiments will be affected, 
among other factors, by our knowledge of Parton Distribution Functions and 
their uncertainties. In this contribution we have reviewed the present status 
of predictions for $t\bar{t}$ and single-top cross-sections evaluated at 
NLO in QCD with different PDF sets and pointed out differences among them, 
trying to elucidate the reasons for these differences. For the case of 
single-top production, we have shown that one important source of difference
among predictions obtained using different PDF sets is the value of the 
$b$-quark mass $m_b$ used by the different collaborations. 

In the second part we briefly discussed PDF-induced correlations between 
parton densities and top cross-sections and between the latter and 
electroweak vector boson production cross-sections at the LHC at 7 TeV. 
These correlation could be useful to define experimental strategies to 
measure the top quark cross-section in a way in which PDF uncertainties are 
reduced.

\acknowledgments
AG would like to thank the organizers, and in particular Fabio Maltoni,
for the kind invitation to participate in a very nice and stimulating 
Workshop and the HEPTOOLS European Network for providing the financial 
support for his participation.


\begin{thebibliography}{0}

\bibitem{Campbell:2006wx}
  \BY{J.~M.~Campbell, J.~W.~Huston \atque W.~J.~Stirling}
  \IN{Rept.\ Prog.\ Phys.}{70}{2007}{89}

\bibitem{Ferrag:2004ca}
  \BY{S.~Ferrag  [ATLAS Collaboration]}
  {\tt arXiv:hep-ph/0407303}

\bibitem{DelDebbio:2007ee}
  \BY{L.~Del Debbio, S.~Forte, J.~I.~Latorre, A.~Piccione \atque J.~Rojo  
    [NNPDF Collaboration]}
  \IN{Jour.\ Hi.\ En.\ Phys.}{0703}{2007}{039}

\bibitem{Ball:2008by}
  \BY{R.~D.~Ball, L.~Del Debbio, S.~Forte, A.~Guffanti, J.~I.~Latorre, 
    A.~Piccione, J.~Rojo \atque M.~Ubiali [NNPDF Collaboration]}
  \IN{Nucl.\ Phys.\  B}{809}{2009}{1}

\bibitem{Ball:2009mk}
  \BY{R.~D.~Ball, L.~Del Debbio, S.~Forte, A.~Guffanti, J.~I.~Latorre, 
    A.~Piccione, J.~Rojo \atque M.~Ubiali [NNPDF Collaboration]}
  \IN{Nucl.\ Phys.\  B}{823}{2009}{195}

\bibitem{Ball:2010de}
  \BY{R.~D.~Ball, L.~Del Debbio, S.~Forte, A.~Guffanti, J.~I.~Latorre, 
    J.~Rojo \atque M.~Ubiali [NNPDF Collaboration]}
  \IN{Nucl.\ Phys.\  B}{838}{2010}{136}

\bibitem{ref:MCFM} 
  \BY{J.~Campbell \atque K.~Ellis}
 \IN{Phys.\ Rev.\  D}{62}{2000}{114012},
  {\tt http://mcfm.fnal.gov}

\bibitem{Nadolsky:2008zw}
  \BY{P.~M.~Nadolsky {\it et al.}}
  \IN{Phys.\ Rev.\  D}{78}{2008}{013004}

\bibitem{Martin:2009iq}
  \BY{A.~D.~Martin, W.~J.~Stirling, R.~S.~Thorne \atque G.~Watt}
  \IN{Eur.\ Phys.\ J.\  C}{63}{2009}{189}

\bibitem{Alekhin:2009ni}
  \BY{S.~Alekhin, J.~Bl\"umlein, S.~Klein and S.~Moch}
  \IN{Phys.\ Rev.\  D}{81}{2010}{014032}

\bibitem{:2009wt}
  \BY{F.~D.~Aaron {\it et al.}  [H1 Collaboration and ZEUS Collaboration]}
  \IN{JHEP}{1001}{2010}{109}

\bibitem{Demartin:2010er}
  \BY{F.~Demartin, S.~Forte, E.~Mariani, J.~Rojo \atque A.~Vicini}
  \IN{Phys.\ Rev.\  D}{82}{2010}{014002}

\bibitem{Ubiali:2010xc}
  \BY{ M.~Ubiali, R.~D.~Ball, L.~Del Debbio, S.~Forte, A.~Guffanti, 
    J.~I.~Latorre \atque J.~Rojo}
  {\tt arXiv:0903.0005}

\bibitem{Campbell:2009ss}
  \BY{J.~M.~Campbell, R.~Frederix, F.~Maltoni \atque F.~Tramontano}
  \IN{Phys.\ Rev.\ Lett.}{102}{2009}{182003}

\end{thebibliography}
\end{document}